\documentclass[conference]{IEEEtran}
\IEEEoverridecommandlockouts
\usepackage{cite}
\usepackage{amsmath,amssymb,amsfonts}
\usepackage{algorithmic}
\usepackage{graphicx}
\usepackage{textcomp}
\usepackage{arydshln}
\usepackage{caption}
\usepackage[justification=centering]{caption}
\usepackage{comment}
\usepackage{url}
\usepackage[symbol]{footmisc}

\usepackage{xcolor}
\def\BibTeX{{\rm B\kern-.05em{\sc i\kern-.025em b}\kern-.08em
    T\kern-.1667em\lower.7ex\hbox{E}\kern-.125emX}}
\begin{document}

\title{Hardware for converting floating-point \\ to the
microscaling (MX) format 
}


\author{\IEEEauthorblockN{Danila Gorodecky}
\IEEEauthorblockA{\textit{Instituto Superior Tecnico,} \\ \textit{Universidade de Lisboa}\\
Lisbon, Portugal \\
danila.gorodecky@gmail.com}
\and
\IEEEauthorblockN{Leonel Sousa}
\IEEEauthorblockA{\textit{Instituto Superior Tecnico,} \\ \textit{Universidade de Lisboa}\\
Lisbon, Portugal \\
las@inesc-id.pt}
}

\maketitle

\begin{abstract}
This paper proposes hardware converters for the microscaling format (MX-format), a reduced representation of floating-point numbers. We present an algorithm and a memory-free hardware model for converting 32 single-precision floating-point numbers to MX-format. The proposed model supports six different types of MX-format: E5M2, E4M3, E3M2, E2M3, E2M1, and INT8. The conversion process consists of three steps: calculating the maximum absolute value among 32 inputs, generating a shared scale, and producing 32 outputs in the selected MX-format type. The hardware converters were implemented in FPGA, and experimental results demonstrate.
\end{abstract}

\begin{IEEEkeywords}
floating-point formats, microscaling format (MX), FPGA.
\end{IEEEkeywords}

\section{Introduction and Background}
The Microscaling (MX) number representation was proposed in 2023 \cite{MX_1,MX_2} through a collaboration between AMD, Arm, Intel, Meta, Microsoft, NVIDIA, and Qualcomm. This format aims to minimize hardware costs and storage requirements via reduced bit-width, while maintaining the efficiency and accuracy needed for machine learning training and inference tasks. MX-formats provide a unified representation for half-precision (FP16), single-precision (FP32), BFloat16, and others formats by using truncated binary formats with a sign bit, up to 5 bits for the exponent, and up to 6 bits for the mantissa. A key feature of the MX-format is its use of a shared scale factor that represents the exponent of the floating-point format.

\begin{figure}[h!]
\centering
\includegraphics[width=250pt]{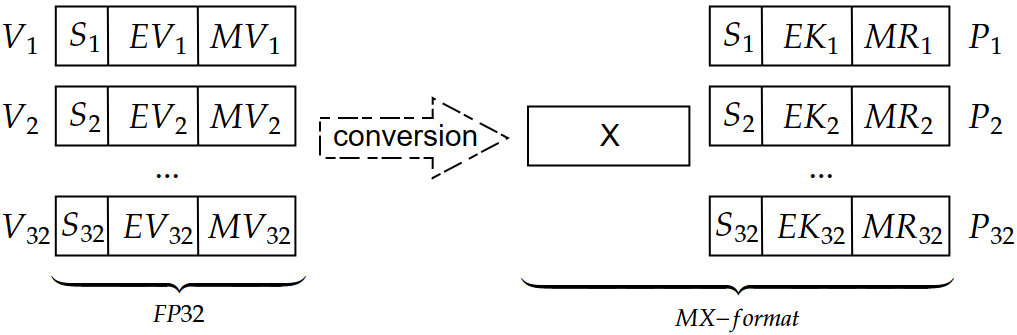}
\caption{The ratio between FP32 and MX-format:\\ \small{$S_1,S_2,\dots,S_{32}$ - sign bits, $EV_1,EV_2,\dots,EV_{32}$ - exponent and $MV_1,MV_2,\dots,MV_{32}$ - mantissa of $V_1,V_2,\dots,V_{32}$,\\
$X$ - shared scale, $EK_1,EK_2,\dots,EK_{32}$ - exponent and $MR_1,MR_2,\dots,MR_{32}$ - mantissa of $P_1,P_2,\dots,P_{32}$, where $K$ and $R$ - nimber of bits of exponent and mantissa, respectively, of MX-format}}\label{MX_common}
\end{figure}

MX-format is the modified floating-point representation with $w$-bit shared scale $X$  (represented in exponent format) and privet $n$ element $P_i$, where $i=1,2,\dots,n$. Fig.~\ref{MX_common} shows the mapping of 32-bits floating point numbers (FP32) $V_1,V_2,\dots, V_{32}$ to MX-format $X,P_1,P_2,\dots, P_{32}$.

MX-format is characterized by the $w$-bit shared scale $X$ and private elements $P_1,P_2,\dots, P_{n}$. This paper considers six types of MX-format. The bit-with of the shared scale depends on the type of MX. The general form of a type is represented as EKMR, where "K" and "R" are the number of bits of exponent (EK on Fig.~\ref{MX_common}) and mantissa (MR on Fig.~\ref{MX_common}) respectively. The sign bit (S on Fig.~\ref{MX_common}) is mentioned by default.

This paper considers the following types of MX-format: E5M2, E4M3, E3M2, E2M3, E2M1, and INT8, which are the ones currently considered for possible standardization. The bit range of the shared scale $X$ for FP32 in the mentioned types of MX-format is 8, i.e. $w=8$. For example, for $n=2$ and E5M2, $V_1,V_2$ in FP32 correspond to two private elements $P_1,P_2$ that consist of 8-bit, i.e. 1 sign bit, 5-bit exponent, and 2-bit mantissa each. The bit-width of the shared scale, sign bit, exponent, and mantissa for the considered formats are shown in Table \ref{tab_1}.   

\setlength{\tabcolsep}{0.9em}
\renewcommand{\arraystretch}{1.4}
\begin{table}[h!]
\centering
\caption{Bit-width of the shared scale ($X$), sign bit $S$, exponent $EK$, and mantissa $MR$ for E5M2, E4M3, E3M2, E2M3, E2M1, and INT8} \label{tab_1}
\begin{tabular}{c|c|c|c|c|}
\cline{2-5} 
                           & X                  & S                  & EK & MR \\ \hline \hline
\multicolumn{1}{|l|}{E5M2} & & & 5  & 2  \\ \cline{1-1} \cline{4-5} 
\multicolumn{1}{|l|}{E4M3} & & & 4  & 3  \\ \cline{1-1} \cline{4-5} 
\multicolumn{1}{|l|}{E3M2} &  {8} &   {1}  & 3  & 2  \\ \cline{1-1} \cline{4-5} 
\multicolumn{1}{|l|}{E2M3} &                    &                    & 2  & 3  \\ \cline{1-1} \cline{4-5} 
\multicolumn{1}{|l|}{E2M1} &                    &                    & 2  & 1  \\ \cline{1-1} \cline{4-5} 
\multicolumn{1}{|l|}{INT8} &                    &                    & 1  & 6  \\ \hline
\end{tabular}
\end{table}

According to the original mechanism \cite{MX_1,MX_2}, the converting consists of two steps:
\begin{enumerate}
    \item calculating $X$;
    \item generating quantized outputs $P_1,P_2,\dots,P_n$. 
\end{enumerate}

The shared scale $X$ can represent NaN, but not infinity. 
$X$ value is treated separately from the private elements. Even when $X$ is NaN, the private elements can be anything (normal number, infinity, and NaN). However, in backward transformation, the resulting values will all be NaN if $X$ is NaN, because multiplication of NaN to a value equals NaN.  

This paper proposes an algorithm and a hardware architecture for converting FP32 to MX formats. The proposed models were implemented on FPGA, with experimental results obtained after synthesis and place-and-route.

In \cite{sam_mel_luk_con}, a converter is described that processes input into blocks of MX values using 'round to nearest even' for FP32 and BFloat16 and focuses on a pipelined (memory-based) design. In contrast, the approach in this paper uses combinational logic (i.e., without memory) and applies standard IEEE 754 rounding.

The rest of the paper is organized as follows. Section II will consider the converting algorithm from FP32 to MX-format and quantization of the considered MX-formats with the examples. Section III will represent the structure of the hardware converter. This will be followed by the results of the synthesis in section IV. Finally, the conclusions are presented in section V.

\section{Converting algorithm FP32 to MX-formats}
The procedure of the transformation reminds the original one, but it splits the first step into two steps, where on the first step the largest power-of-two among all inputs is calculated, and then the largest value is tis defined, thus transformed to $X$ (see Table~\ref{tab_2}). As a result, the procedure of the transformation is the following:
\begin{enumerate}
    \item identifying the largest power of two, denoted as $max(|V_i|)$, among all inputs $V_1, V_2, \dots, V_n$. The largest power of two is determined by examining the exponents $EV_1, EV_2, \dots, EV_n$, so that $max(|V_i|)=max |\{EV_i\} |$. This maximum value is calculated by comparing pairs of values in a hierarchical manner: first, $EV_1$ with $EV_2$, $EV_3$ with $EV_4$, and so on, up to $EV_{31}$ with $EV_{32}$. At each level, the larger value in each pair is compared with the larger value from the adjacent pair. This process continues until, on the final (fifth) step, the largest value among all 32 inputs is determined;
    \item the transformation of $max(|EV_i|)$ into $X$ is achieved by converting $max(|EV_i|)$ to the order $O_{max}$, then subtracting the value of the largest order of the considered data type $EK$ in the MX-format from $O_{max}$. For example, if $max(|V_i|)= 0 \underbrace{10101011}_\text{8-bit exponent}011\dots$, then $max(|EV_i|)=10101011$, which corresponds to $EV_i = 171$ and $O_{max}(EV)=44$. The value of the largest order of E5M2 is 15, with $E5 = 11110$; thus $X=O_{max} - E5 = 44-15 = 29$. The ratio of exponent and order of numbers in format FP32 to the shared scale $X$ for E5M2, E4M3, E3M2, E2M3, E2M1, and INT8 is shown in Table \ref{tab_2};
    \item the calculation and quantization of $P_1,P_2,\dots,P_n$ involve rounding $r+1$ input bits to $r$ output bits. For example, the three most significant bits of the mantissa in FP32 are rounded to two bits of the mantissa in the E5M2 format. The quantization details for E5M2, E4M3, E3M2, E2M3, and E2M1 formats are presented in Tables \ref{tab_3}, \ref{tab_4}, \ref{tab_5}, \ref{tab_6}, and \ref{tab_7}, respectively. The quantization of INT8 is not considered here due to the extensive 65-line table required.
    
\end{enumerate}

\setlength{\tabcolsep}{0.75em}
\renewcommand{\arraystretch}{1.7}
\begin{table*}[h]
  \centering
  \caption{The ratio of exponent and order of numbers in format FP32 to the shared scale $X$ of E5M2, E4M3, E3M2, E2M3 E2M1, and INT8} \label{tab_2}
\begin{tabular}{|c|c|c:c:cccccccccccc:c:c|}
\hline
{FP32}                                                & order    & 0    & 1    & \multicolumn{1}{c:}{2}    & \multicolumn{1}{c:}{3}    & \multicolumn{1}{c:}{4}    & \multicolumn{1}{c:}{5} & \multicolumn{1}{c:}{6} & \multicolumn{1}{c:}{7} & \multicolumn{1}{c:}{8}    & \multicolumn{1}{c:}{...} & \multicolumn{1}{c:}{15} & \multicolumn{1}{c:}{16}   & \multicolumn{1}{c:}{17}   & \dots & 254 & 255                                       \\ \cdashline{2-18} 
                                                                     & exponent & -126 & -126 & \multicolumn{1}{c:}{-125} & \multicolumn{11}{c:}{\dots}                                                                                                                                                                                                                                                       & 127 & $\infty$ / NaN \\ \hline \hline
{E5M2}                                                & order    & 0    & 0    & \multicolumn{8}{c:}{...}                                                                                                                                                                                            & \multicolumn{1}{c:}{0}  & \multicolumn{1}{c:}{1}    & \multicolumn{1}{c:}{2}    & ... & 239 & 240                                       \\ \cdashline{2-18} 
                                                                     & exponent & 0    & 0    & \multicolumn{8}{c:}{...}                                                                                                                                                                                            & \multicolumn{1}{c:}{0}  & \multicolumn{1}{c:}{-126} & \multicolumn{1}{c:}{-125} & ... & 112 & 
                                                                     NaN \\ \hline
{E4M3}                                                & order    & 0    & 0    & \multicolumn{5}{c:}{...}                                                                                                            & \multicolumn{1}{c:}{0} & \multicolumn{1}{c:}{1}    & \multicolumn{5}{c:}{...}                                                                                         & 247 & 248                                       \\ \cdashline{2-18} 
                                                                     & exponent & 0    & 0    & \multicolumn{5}{c:}{...}                                                                                                            & \multicolumn{1}{c:}{0} & \multicolumn{1}{c:}{-126} & \multicolumn{5}{c:}{...}                                                                                         & 120 & 
                                                                     NaN \\ \hline
{E3M2}                                                & order    & 0    & 0    & \multicolumn{1}{c:}{0}    & \multicolumn{1}{c:}{0}    & \multicolumn{1}{c:}{1}    & \multicolumn{9}{c:}{...}                                                                                                                                                                                                & 251 & 252                                       \\ \cdashline{2-18} 
                                                                     & exponent & 0    & 0    & \multicolumn{1}{c:}{0}    & \multicolumn{1}{c:}{0}    & \multicolumn{1}{c:}{-126} & \multicolumn{9}{c:}{...}                                                                                                                                                                                                & 124 & 
                                                                     NaN \\ \hline
{\begin{tabular}[c]{@{}l@{}}E2M3\end{tabular}} & order    & 0    & 0    & \multicolumn{1}{c:}{1}    & \multicolumn{1}{c:}{2}    & \multicolumn{10}{c:}{...}                                                                                                                                                                                                                           & 253 & 254                                       \\ \cdashline{2-18} 
{\begin{tabular}[c]{@{}l@{}}E2M1\end{tabular}}                                                                     & exponent & 0    & 0    & \multicolumn{1}{c:}{-126} & \multicolumn{1}{c:}{-125} & \multicolumn{10}{c:}{...}                                                                                                                                                                                                                           & 126 & 
/ NaN \\ \hline
{INT8}                                                & order    & 0    & 1    & \multicolumn{1}{c:}{2}    & \multicolumn{11}{c:}{...}                                                                                                                                                                                                                                                       & 254 & 255                                       \\ \cdashline{2-18} 
                                                                     & exponent & 0    & -126 & \multicolumn{1}{c:}{-125} & \multicolumn{11}{c:}{...}                                                                                                                                                                                                                                                       & 127 & 
                                                                     NaN \\ \hline
\end{tabular}
\end{table*}

\setlength{\tabcolsep}{0.25em}
\renewcommand{\arraystretch}{1.7}

\begin{table}[]
  \centering
  \caption{Quantization of E5M2}\label{tab_3}
\begin{tabular}{|c||ccc|cc|} 
\hline
\multicolumn{1}{|c||}{E5}                                                                    &
\multicolumn{3}{|c|}{$V[23:21]$}                                                                    & \multicolumn{2}{c|}{$M2$}      \\ \hline
E[5] E[4] E[3] E[2] E[1] &$V[23]$     & $V[22]$ & $V[21]$     & $M[2]$ & $M[1]$ \\ \hline
\multicolumn{1}{|c||}{$E[5]\&E[4]\&E[3]\&E[2]\&\overline{E[1]}\neq 1$}& 0  & 0  & 0  & 0   & 0   \\ \hline
\multicolumn{1}{|c||}{$E[5]\&E[4]\&E[3]\&E[2]\&\overline{E[1]}\neq 1$}& 0  & 0  & 1  & 0   & 1     \\ \hline
\multicolumn{1}{|c||}{$E[5]\&E[4]\&E[3]\&E[2]\&\overline{E[1]}\neq 1$}& 0  & 1  & 0  & 1   & 1     \\ \hline
\multicolumn{1}{|c||}{$E[5]\&E[4]\&E[3]\&E[2]\&\overline{E[1]}\neq 1$}& 0  & 1  & 1  & 1   & 0     \\ \hline
\multicolumn{1}{|c||}{$E[5]\&E[4]\&E[3]\&E[2]\&\overline{E[1]}\neq 1$}& 1  & 0  & 0  & 1   & 0     \\ \hline
\multicolumn{1}{|c||}{$E[5]\&E[4]\&E[3]\&E[2]\&\overline{E[1]}\neq 1$}& 1  & 0  & 1  & 1   & 1     \\ \hline
\multicolumn{1}{|c||}{$E[5]\&E[4]\&E[3]\&E[2]\&\overline{E[1]}\neq 1$}& 1  & 1  & 0  & 1   & 1     \\ \hline
\multicolumn{1}{|c||}{$E[5]\&E[4]\&E[3]\&E[2]\&\overline{E[1]}\neq 1$}& 1  & 1  & 1  & 0   & 0     \\ \hline
\multicolumn{1}{|c||}{$E[5]\&E[4]\&E[3]\&E[2]\&\overline{E[1]}=1$}    & 1  & 1  & 1  & 1   & 1     \\ \hline
\end{tabular}
\end{table}

\setlength{\tabcolsep}{0.12em}
\renewcommand{\arraystretch}{1.7}

\begin{table}[]
  \centering
  \caption{ Quantization of E4M3} \label{tab_4}
\begin{tabular}{|c||cccc|ccc|} 
\hline
\multicolumn{1}{|c||}{E4}                                                                    &
\multicolumn{4}{|c|}{$V[23:20]$}                                                                    & \multicolumn{3}{c|}{$M3$}      \\ \hline
E[4] E[3] E[2] E[1] &$V[23]$  & $V[22]$ & $V[21]$ & $V[20]$   & $M[3]$  & $M[2]$ & $M[1]$ \\ \hline
\multicolumn{1}{|c||}{$E[4]\&E[3]\&E[2]\&\overline{E[1]}\neq 1$}& 0 & 0   & 0   & 0   & 0   & 0  & 0     \\ \hline
\multicolumn{1}{|c||}{$E[4]\&E[3]\&E[2]\&\overline{E[1]}\neq 1$}& 0 & 0   & 0   & 1   & 0   & 0  & 1     \\ \hline
\multicolumn{1}{|c||}{$E[4]\&E[3]\&E[2]\&\overline{E[1]}\neq 1$}& 0 & 0   & 1   & 0   & 0   & 0  & 1     \\ \hline
\multicolumn{1}{|c||}{$E[4]\&E[3]\&E[2]\&\overline{E[1]}\neq 1$}& 0 & 0   & 1   & 1   & 0   & 1  & 0     \\ \hline
\multicolumn{1}{|c||}{$E[4]\&E[3]\&E[2]\&\overline{E[1]}\neq 1$}& 0 & 1   & 0   & 0   & 0   & 1  & 0     \\ \hline
\multicolumn{1}{|c||}{$E[4]\&E[3]\&E[2]\&\overline{E[1]}\neq 1$}& 0 & 1   & 0   & 1   & 0   & 1  & 1     \\ \hline
\multicolumn{1}{|c||}{$E[4]\&E[3]\&E[2]\&\overline{E[1]}\neq 1$}& 0 & 1   & 1   & 0   & 0   & 1  & 1     \\ \hline
\multicolumn{1}{|c||}{$E[4]\&E[3]\&E[2]\&\overline{E[1]}\neq 1$}& 0 & 1   & 1   & 1   & 1   & 0  & 0     \\ \hline
\multicolumn{1}{|c||}{$E[4]\&E[3]\&E[2]\&\overline{E[1]}\neq 1$}& 1 & 0   & 0   & 0   & 1   & 0  & 0     \\ \hline
\multicolumn{1}{|c||}{$E[4]\&E[3]\&E[2]\&\overline{E[1]}\neq 1$}& 1 & 0   & 0   & 1   & 1   & 0  & 1     \\ \hline
\multicolumn{1}{|c||}{$E[4]\&E[3]\&E[2]\&\overline{E[1]}\neq 1$}& 1 & 0   & 1   & 0   & 1   & 0  & 1     \\ \hline
\multicolumn{1}{|c||}{$E[4]\&E[3]\&E[2]\&\overline{E[1]}\neq 1$}& 1 & 0   & 1   & 1   & 1   & 1  & 0     \\ \hline
\multicolumn{1}{|c||}{$E[4]\&E[3]\&E[2]\&\overline{E[1]}\neq 1$}& 1 & 1   & 0   & 0   & 1   & 1  & 0     \\ \hline
\multicolumn{1}{|c||}{$E[4]\&E[3]\&E[2]\&\overline{E[1]}\neq 1$}& 1 & 1   & 0   & 1   & 1   & 1  & 1     \\ \hline
\multicolumn{1}{|c||}{$E[4]\&E[3]\&E[2]\&\overline{E[1]}\neq 1$}& 1 & 1   & 1   & 0   & 1   & 1  & 1     \\ \hline
\multicolumn{1}{|c||}{$E[4]\&E[3]\&E[2]\&\overline{E[1]}\neq 1$}& 1 & 1   & 1   & 1   & 0   & 0  & 0     \\ \hline
\multicolumn{1}{|c||}{$E[4]\&E[3]\&E[2]\&\overline{E[1]}= 1$}& 1 & 1   & 1   & 1   & 1   & 1  & 1     \\ \hline
\end{tabular}
\end{table}

\setlength{\tabcolsep}{0.29em}
\renewcommand{\arraystretch}{1.7}

\begin{table}[]
  \centering
  \caption{Quantization of E3M2} \label{tab_5}
\begin{tabular}{|c||ccc|cc|} 
\hline
\multicolumn{1}{|c||}{E3}                                                                    &
\multicolumn{3}{|c|}{$V[23:21]$}                                                                    & \multicolumn{2}{c|}{$M2$}      \\ \hline
E[3] E[2] E[1] &$V[23]$     & $V[22]$ & $V[21]$     & $M[2]$ & $M[1]$ \\ \hline
\multicolumn{1}{|c||}{$E[3]\&E[2]\&\overline{E[1]}\neq 1$}& 0  & 0  & 0  & 0   & 0   \\ \hline
\multicolumn{1}{|c||}{$E[3]\&E[2]\&\overline{E[1]}\neq 1$}& 0  & 0  & 1  & 0   & 1     \\ \hline
\multicolumn{1}{|c||}{$E[3]\&E[2]\&\overline{E[1]}\neq 1$}& 0  & 1  & 0  & 1   & 1     \\ \hline
\multicolumn{1}{|c||}{$E[3]\&E[2]\&\overline{E[1]}\neq 1$}& 0  & 1  & 1  & 1   & 0     \\ \hline
\multicolumn{1}{|c||}{$E[3]\&E[2]\&\overline{E[1]}\neq 1$}& 1  & 0  & 0  & 1   & 0     \\ \hline
\multicolumn{1}{|c||}{$E[3]\&E[2]\&\overline{E[1]}\neq 1$}& 1  & 0  & 1  & 1   & 1     \\ \hline
\multicolumn{1}{|c||}{$E[3]\&E[2]\&\overline{E[1]}\neq 1$}& 1  & 1  & 0  & 1   & 1     \\ \hline
\multicolumn{1}{|c||}{$E[3]\&E[2]\&\overline{E[1]}\neq 1$}& 1  & 1  & 1  & 0   & 0     \\ \hline
\multicolumn{1}{|c||}{$E[3]\&E[2]\&\overline{E[1]}=1$}    & 1  & 1  & 1  & 1   & 1     \\ \hline
\end{tabular}
\end{table}

\setlength{\tabcolsep}{0.12em}
\renewcommand{\arraystretch}{1.7}

\begin{table}[]
  \centering
  \caption{Quantization of E2M3} \label{tab_6}
\begin{tabular}{|c||cccc|ccc|} 
\hline
\multicolumn{1}{|c||}{E2}                                                                    &
\multicolumn{4}{|c|}{$V[23:20]$}                                                                    & \multicolumn{3}{c|}{$M2$}      \\ \hline
E[2] E[1] &$V[23]$  & $V[22]$ & $V[21]$ & $V[20]$   & $M[3]$  & $M[2]$ & $M[1]$ \\ \hline
\multicolumn{1}{|c||}{$E[2]\&\overline{E[1]}\neq 1$}& 0 & 0   & 0   & 0   & 0   & 0  & 0     \\ \hline
\multicolumn{1}{|c||}{$E[2]\&\overline{E[1]}\neq 1$}& 0 & 0   & 0   & 1   & 0   & 0  & 1     \\ \hline
\multicolumn{1}{|c||}{$E[2]\&\overline{E[1]}\neq 1$}& 0 & 0   & 1   & 0   & 0   & 0  & 1     \\ \hline
\multicolumn{1}{|c||}{$E[2]\&\overline{E[1]}\neq 1$}& 0 & 0   & 1   & 1   & 0   & 1  & 0     \\ \hline
\multicolumn{1}{|c||}{$E[2]\&\overline{E[1]}\neq 1$}& 0 & 1   & 0   & 0   & 0   & 1  & 0     \\ \hline
\multicolumn{1}{|c||}{$E[2]\&\overline{E[1]}\neq 1$}& 0 & 1   & 0   & 1   & 0   & 1  & 1     \\ \hline
\multicolumn{1}{|c||}{$E[2]\&\overline{E[1]}\neq 1$}& 0 & 1   & 1   & 0   & 0   & 1  & 1     \\ \hline
\multicolumn{1}{|c||}{$E[2]\&\overline{E[1]}\neq 1$}& 0 & 1   & 1   & 1   & 1   & 0  & 0     \\ \hline
\multicolumn{1}{|c||}{$E[2]\&\overline{E[1]}\neq 1$}& 1 & 0   & 0   & 0   & 1   & 0  & 0     \\ \hline
\multicolumn{1}{|c||}{$E[2]\&\overline{E[1]}\neq 1$}& 1 & 0   & 0   & 1   & 1   & 0  & 1     \\ \hline
\multicolumn{1}{|c||}{$E[2]\&\overline{E[1]}\neq 1$}& 1 & 0   & 1   & 0   & 1   & 0  & 1     \\ \hline
\multicolumn{1}{|c||}{$E[2]\&\overline{E[1]}\neq 1$}& 1 & 0   & 1   & 1   & 1   & 1  & 0     \\ \hline
\multicolumn{1}{|c||}{$E[2]\&\overline{E[1]}\neq 1$}& 1 & 1   & 0   & 0   & 1   & 1  & 0     \\ \hline
\multicolumn{1}{|c||}{$E[2]\&\overline{E[1]}\neq 1$}& 1 & 1   & 0   & 1   & 1   & 1  & 1     \\ \hline
\multicolumn{1}{|c||}{$E[2]\&\overline{E[1]}\neq 1$}& 1 & 1   & 1   & 0   & 1   & 1  & 1     \\ \hline
\multicolumn{1}{|c||}{$E[2]\&\overline{E[1]}\neq 1$}& 1 & 1   & 1   & 1   & 0   & 0  & 0     \\ \hline
\multicolumn{1}{|c||}{$E[2]\&\overline{E[1]}= 1$}& 1 & 1   & 1   & 1   & 1   & 1  & 1     \\ \hline
\end{tabular}
\end{table}

\setlength{\tabcolsep}{0.2em}
\renewcommand{\arraystretch}{1.7}

\begin{table}[]
  \centering
  \caption{Quantization of E2M1} \label{tab_7}
\begin{tabular}{|c||cc|c|} 
\hline
\multicolumn{1}{|c||}{E2}                                                                    &
\multicolumn{2}{|c|}{$V[23:22]$}                                                                    & \multicolumn{1}{c|}{$M1$}      \\ \hline
E[2] E[1] &$V[23]$     & $V[22]$    &  $M[1]$ \\ \hline
\multicolumn{1}{|c||}{$E[2]\&\overline{E[1]}\neq 1$}& 0  & 0  & 0     \\ \hline
\multicolumn{1}{|c||}{$E[2]\&\overline{E[1]}\neq 1$}& 0  & 1  & 1       \\ \hline
\multicolumn{1}{|c||}{$E[2]\&\overline{E[1]}\neq 1$}& 1  & 0  & 0       \\ \hline
\multicolumn{1}{|c||}{$E[2]\&\overline{E[1]}\neq 1$}& 1  & 1  & 0      \\ \hline
\multicolumn{1}{|c||}{$E[2]\&\overline{E[1]}=1$}    & 1  & 1  & 1       \\ \hline
\end{tabular}
\end{table}



The shared scale $X$ can represent NaN, but it cannot represent infinity. The value of $X$ is handled separately from the private elements. Even if $X$ is NaN, the private elements can still be normal numbers, infinities, or NaNs. However, if $X$ is NaN, all resulting private values will also be NaN.

\section{Architecture of the Proposed Converter}
\begin{figure*}
\centering
    \includegraphics[width=510pt]{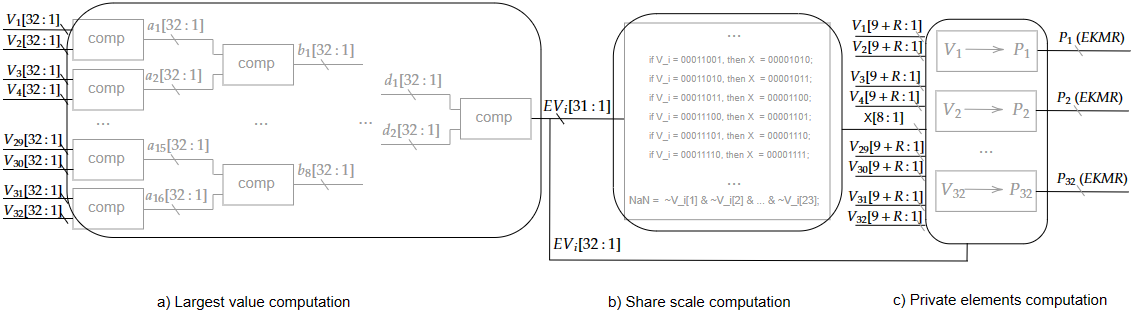}
    \caption{Architecture of the converter}
    {\small{a) The largest value calculation $max(|V_i|)=E{V_i}$, where $a_1, a_2, \cdots, a_{16}$ are outputs of the first layer of the comparison, $b_1, b_2, \cdots, b_8$ are outputs of the second layer of the comparison, $c_1, c_2, c_3, c_4$ are outputs of the third layer of the comparison, $d_1, d_2$ are outputs of the fourth layer of the comparison \\}}
    {\small{b) The shared scale computation block transforms 31-bit of the value $E{V_i}$ to the shared scale value $X$ \\}}
    {\small{c) The private elements computation block processes the shred scale $X$ and $9+R$ most significant bits of FP32 inputs, where $R$ is the bit-range of the mantissa of the appropriate types of MX-format. The private elements $P_1,P_2,\dots,P_{32}$ consist of $1 + K + R$-bits, where $K$ is bit-range of the exponent of the appropriate types of MX-format}}
    \label{com_arch}
\end{figure*}

The proposed procedure includes three steps (see Fig.~\ref{com_arch}) by split calculating $X$ into two steps, as mentioned in the previous section\footnote[1]{Verilog-files for the transformation of FP32 to six types of MX: \\ \url{https://github.com/ZeboZebo702/MX-formats/tree/main/FP32_to_MX}}:
\begin{enumerate}
    \item defining the largest power-of-two ($max(|EV_i|)$) among all inputs $V_1,V_2,\dots,V_n$;
    \item transformation of $max(|EV_i|)$ into $X$;
    \item calculating and quantization of $P_1,P_2,\dots,P_n$. 
\end{enumerate}

\subsection{Defining the largest among $V_1,V_2,\dots,V_n$ ($max(|EV_i|)$)}
\renewcommand{\thefootnote}{\arabic{footnote}}
For calculating the largest value $max(|EV_i|)$ among $V_1,V_2,\dots,V_n$ is processed by comparing pairs of values in a hierarchical manner in five steps as pictured on Fig.~\ref{com_arch}. The core of the comparison is the module "comp" with two 32-bit inputs and one 32-bit output.

The module "comp" calculates the following cases:
\begin{itemize}
    \item if both exponents of FP32 $x$ and $y$ are equal 8-bit of $11111111$, then 32-output equals 32-bit of $0$; 
    \item if one of exponents of FP32 $x$ are equal 8-bit of $11111111$, (i.e. FP32 is "0" or $\pm \infty$) and the other $y$ equals another value, then 32-output equals $y$; 
    \item if both exponents of FP32 $x$ and $y$ are not equal 8-bit of $11111111$, (i.e. FP32 are not "0" or $\pm \infty$), then 32-output equals the largest of $x$ and $y$.
\end{itemize}

\textit{Example. Part 1.}
Consider four FP32 inputs:
\begin{enumerate}
    \item $V_1 =\underbrace{0}_\text{sign} \underbrace{10101011}_\text{8-bit exponent} \underbrace{011\cdots}_\text{23-bit mantissa}$, 
    \item $V_2 =\underbrace{0}_\text{sign} \underbrace{10101000}_\text{8-bit exponent} \underbrace{110\cdots}_\text{23-bit mantissa}$, 
    \item $V_3 =\underbrace{0}_\text{sign} \underbrace{00101011}_\text{8-bit exponent} \underbrace{001\cdots}_\text{23-bit mantissa}$,
    \item $V_4= \underbrace{1}_\text{sign} \underbrace{10001111}_\text{8-bit exponent} \underbrace{010\cdots}_\text{23-bit mantissa}$.
\end{enumerate}
The result of the comparison of absolute values of $V_1$, $V_2$, $V_3$, $V_4$ gives $V_1$ as largest power-of-two result $max(|EV_i|) = E{V_1} = 10101011$. 

\subsection{Transformation of $max(|EV_i|)$ into $X$}
The transformation step consists of the module "div", which transforms the largest value $max(|EV_i|)$ to the shared scale $X$ (central part of Fig.~\ref{com_arch} b)). The 31-bit input of this module represented as the least significant bits of FP32 (all bits of FP32 without the "sign" bit) is transformed to 8-bit output $X$. 8-bit exponent $E{V_i}$ ($ [31:24 ]$ of $E$ bit of $E{V_i}$) is encoded to intermediate value $X_{temp}$ according to the rule:
$$
X_{temp} =     
\begin{cases}
  E{V_i} - 2^{K-1}, \; \textbf{if} \; E{V_i} > 2^{K-1}-1;\\    
  0, \textbf{in other cases},    
\end{cases}
$$
where $K$ is the bit-with of the exponent in MX-format type.

For example, if $E{V_i}= 00001011$, then for $E5M2$, where $K=5$, $X_{temp} = 00000000$; if $E{V_i}= 11100010$, then for $E5M2$, where $K=5$, $X_{temp} = 11010011$.

The 1-bit value $NaN$ is calculated on this step. 
$$
NaN = \overline{V_i[1]} \ \& \ \overline{V_i[2]} \ \& \ \dots \ \& \ \overline{V_i[23]},
$$
where $V_i[1],V_i[2],\dots,V_i[23]$ are 23 mantissa bits of $E{V_i}$.

The output $X$ of the module "div" is calculated in the following way:
\begin{itemize}
    \item $X_{temp} = 11110000$ for E5M2, $X_{temp} = 11111000$ for E4M2, $X_{temp} = 11111100$ for E3M2, $X_{temp} = 1111110$ for E2M3 and E2M1, (i.e. $E{V_i} = 11111111$), and $NAN=0$, then $X=11111111$. i.e. $X$ is considered to the value NAN; 
    \item $X_{temp} = 11110000$ for E5M2, $X_{temp} = 11111000$ for E4M2, $X_{temp} = 11111100$ for E3M2, $X_{temp} = 1111110$ for E2M3 and E2M1, (i.e. $E{V_i} \neq 11111111$), and $NAN=1$, then $X=11111110$. i.e. $X$ is considered as infinity without sign identification;
    \item $X_{temp} \neq 11110000$ for E5M2, $X_{temp} \neq 11111000$ for E4M2, $X_{temp} \neq 11111100$ for E3M2, $X_{temp} \neq 1111110$ for E2M3 and E2M1, (i.e. $E{V_i} \neq 11111111$), then $X=X_{temp}$.
\end{itemize}

\textit{Example. Part 2.}
As defined above, $max(|V_i|) = E{V_1} = 10101011$ and $E{V_1} = 10101011 > 2^{5-1}-1 = 1111$. Hence, $X_{temp}= 10101011 - 1111 = 10011100$. $X_{temp} \neq 11110000$, then $X = X_{temp}= 10011100$.

\subsection{Calculating and quantization of $P_1,P_2,\dots,P_n$}
In this section, we provide rules of the quantization of the considered types of MX-format according to \cite{MX_1,MX_2}. Hence, the third step of the conversion scales and quantilizas $V_i$ to the appropriate type of MX-format in the module "$P_i$" (right part of Fig.~\ref{com_arch} c)). The module is equipped with 8-bit input $X$, 
\begin{itemize}
    \item 12-bit input, where the most significant bit of FP32 $V_i$, $[11:4]$ bits are the exponent of $V_i$, and $[3:1]$ bits correspond to $[23:21]$ and 8-bit output $P_i$ for E5M2 MX-format;
    \item 13-bit input, where the most significant bit of FP32 $V_i$, $[12:5]$ bits are the exponent of $V_i$, and $[4:1]$ bits correspond to $[23:20]$ and 8-bit output $P_i$ for E4M3 MX-format;
    \item 12-bit input, where the most significant bit of FP32 $V_i$, $[11:4]$ bits are the exponent of $V_i$, and $[3:1]$ bits correspond to $[23:21]$  and 6-bit output $P_i$ for E3M2 MX-format;
    \item 13-bit input, where the most significant bit of FP32 $V_i$, $[12:5]$ bits are the exponent of $V_i$, and $[4:1]$ bits correspond to $[23:20]$ and 6-bit output $P_i$ for E2M3 MX-format;
    \item 11-bit input, where the most significant bit of FP32 $V_i$, $[10:3]$ bits are the exponent of $V_i$, and $[2:1]$ bits correspond to $[23:22]$ and 4-bit output $P_i$ for E2M1 MX-format.
\end{itemize}

The result $P_i$ calculates according to the following conditions:
\begin{itemize}
    \item if $X = 11111111$ (i.e. $X$ is NaN), then
        \begin{itemize}
            \item $P_i = V[12] 11111 10$ for E5M2;
            \item $P_i = V[12] 1111 110$ for E4M3;
            \item $P_i = V[12] 111 10$ for E3M2;
            \item $P_i = V[12] 11 110$ for E2M3;
            \item $P_i = V[12] 11 1$ for E2M1;
        \end{itemize}
    \item if $X = 11111110$ (i.e. $X$ is infinity), then 
        \begin{itemize}
            \item $P_i = V[12] 11111 00$ for E5M2;
            \item $P_i = V[12] 1111 000$ for E4M3;
            \item $P_i = V[12] 111 00$ for E3M2;
            \item $P_i = V[12] 11 00$ for E2M3;
            \item $P_i = V[12] 11 0$ for E2M1;
        \end{itemize}
    \item if $X < 11110000$, then $EK = X + 2^{K-1} - 1 \pm E$, where $E$ is the exponent of $V_i$ and $V_i[32] = 0$ corresponds to "$+$" and $V_i[32] = 1$ corresponds to "$-$" in $\pm E$; then after if $EK > 2^{K}$, then $EK: = 0$ (with the same number of bits as $K$) and $MR:=0$ (with the same number of bits as $R$)), else $EK:=2^K-2 - EK$, then
    
         \begin{itemize}
            \item if $EK = 2^K - 1$ and 
                \begin{itemize} 
                    \item $V_i[3:1] = 111$, then $P_i = V[12]*1111011$ for E5M2, i.e. there is no quatilization;
                    \item $V_i[4:1] = 1111$, then $P_i = V[12]*1110111$ for E4M3, i.e. there is no quatilization;
                    \item $V_i[3:1] = 111$, then $P_i = V[12]*11011$ for E3M2, i.e. there is no quatilization;
                    \item $V_i[4:1] = 1111$, then $P_i = V[12]*10111$ for E2M3, i.e. there is no quatilization;
                    \item $V_i[1] = 1$, then $P_i = V[12]*101$ for E2M1, i.e. there is no quatilization;
                \end{itemize}
            \item if $EK \neq 2^K - 1$ and 
                \begin{itemize} 
                    \item $V_i[3:1] = 111$, then $P_i = V[12]*(EK + 1) * 00$ for E5M2;
                    \item $V_i[4:1] = 1111$, then $P_i = V[12]*(EK + 1) * 000$ for E4M3;
                    \item $V_i[3:1] = 111$, then $P_i = V[12]*(EK + 1) * 00$ for E3M2;
                    \item $V_i[4:1] = 1111$, then $P_i = V[12]*(EK + 1) * 000$ for E2M3;
                    \item $V_i[1] = 1$, then $P_i = V[12]*(EK + 1) * 0$ for E2M1;
                \end{itemize}
            \end{itemize}
            \begin{itemize}
                \item if $V_i[3:1] = 110$ or $V_i[3:1] = 101$, then $P_i = V[12]*EK * 11$ for E5M2;
                \item if $V_i[3:1] = 010$ or $V_i[3:1] = 001$, then $P_i = V[12]*EK * 01$ for E5M2;
                \item if $V_i[3:1] = 011$, then $P_i = V[12]*EK * 10$ for E5M2;
            \end{itemize}          
            \begin{itemize}
                \item if $V_i[4:1] = 1111$ or$V_i[4:1] = 1110$ or $V_i[4:1] = 1101$, then $P_i = V[12]*EK * 111$ for E4M3;
                \item if $V_i[4:1] = 1011$ or $V_i[4:1] = 1100$, then $P_i = V[12]*EK * 110$ for E4M3;
                \item if $V_i[4:1] = 1001$ or $V_i[4:1] = 1010$, then $P_i = V[12]*EK * 101$ for E4M3;
                \item if $V_i[4:1] = 0111$ or $V_i[4:1] = 1000$, then $P_i = V[12]*EK * 100$ for E4M3;
                \item if $V_i[4:1] = 0101$ or $V_i[4:1] = 0110$, then $P_i = V[12]*EK * 011$ for E4M3;
                \item if $V_i[4:1] = 0010$ or $V_i[4:1] = 0011$ or $V_i[4:1] = 0100$, then $P_i = V[12]*EK * 010$ for E4M3;
                \item if $V_i[4:1] = 0001$, then $P_i = V[12]*EK * 001$ for E4M3;
            \end{itemize}          
            \begin{itemize}
                \item if $V_i[3:1] = 110$ or $V_i[3:1] = 101$, then $P_i = V[12]*EK * 11$ for E3M2;
                \item if $V_i[3:1] = 010$ or $V_i[3:1] = 001$, then $P_i = V[12]*EK * 01$ for E3M2;
                \item if $V_i[3:1] = 011$, then $P_i = V[12]*EK * 10$ for E3M2;
            \end{itemize}          
            \begin{itemize}
                \item if $V_i[4:1] = 1111$ or$V_i[4:1] = 1110$ or $V_i[4:1] = 1101$, then $P_i = V[12]*EK * 111$ for E2M3;
                \item if $V_i[4:1] = 1011$ or $V_i[4:1] = 1100$, then $P_i = V[12]*EK * 110$ for E2M3;
                \item if $V_i[4:1] = 1001$ or $V_i[4:1] = 1010$, then $P_i = V[12]*EK * 101$ for E2M3;
                \item if $V_i[4:1] = 0111$ or $V_i[4:1] = 1000$, then $P_i = V[12]*EK * 100$ for E2M3;
                \item if $V_i[4:1] = 0101$ or $V_i[4:1] = 0110$, then $P_i = V[12]*EK * 011$ for E2M3;
                \item if $V_i[4:1] = 0010$ or $V_i[4:1] = 0011$ or $V_i[4:1] = 0100$, then $P_i = V[12]*EK * 010$ for E4M3;
                \item if $V_i[4:1] = 0001$, then $P_i = V[12]*EK * 001$ for E4M3;
            \end{itemize}
            \begin{itemize}
                \item if $V_i[3:1] = 10$ or $V_i[3:1] = 01$ or or $V_i[3:1] = 11$, then $P_i = V[12]*EK * 1$ for E2M1.
            \end{itemize}
\end{itemize}

\textit{Example. Part 3.}
Convert four input FP32 to E5M2. As far as $X < 11110000$, then 
\begin{enumerate}
    \item for $V_1 = 0\underbrace{10101011}_\text{E} 011\dots$: $EK = 156 + 2^{5-1} - 1 - 171$, then $EK: = 2^K - 2 - 0 = 30$; $V_1[23:21]=011$, then $MK=10$. Hence, $P_1=V[32]*11110*10=01111010$;
    \item for $V_2 = 0\underbrace{10101000}_\text{E} 110\dots$: $EK = 156 + 2^{5-1} - 1 - 168$, then $EK: = 2^K - 2 - 3 = 27$; $V_2[23:21]=110$, then $MK=11$. Hence, $P_2=V[32]*11011*11=01101111$;
    \item for $V_3 = 0\underbrace{00101011}_\text{E} 001\dots$: $EK = 156 + 2^{5-1} - 1 - 43$. As far as $EK > 2^5$, then $EK: = 0$ and $MR: = 0$. Hence, $P_3=V[32]*00000*00=00000000$;
    \item for $V_4 = 1\underbrace{10001111}_\text{E} 001\dots$: $EK = 156 + 2^{5-1} - 1 + 143$. As far as $EK > 2^5$, then $EK: = 0$ and $MR: = 0$. Hence, $P_4=V[32]*00000*00=10000000$.
\end{enumerate}

\section{Experimental results}
The paper considers the conversion of 32-bit of FP32 to 32 EKMR of MX-format. It means that converter has 
\begin{itemize}
    \item $\underbrace{32\cdot32}_\text{inputs of $V_1,V_2\dots , V_{32}$} + \underbrace{8}_\text{output $X$} + \underbrace{32\cdot8}_\text{output $P_1,P_2\dots , P_{32}$} = 1288$ input/output blocks for E5M2; 
    \item $\underbrace{32\cdot32}_\text{inputs of $V_1,V_2\dots , V_{32}$} + \underbrace{8}_\text{output $X$} + \underbrace{32\cdot8}_\text{output $P_1,P_2\cdot , P_{32}$} = 1288$ input/output blocks for E4M3;
    \item $\underbrace{32\cdot32}_\text{inputs of $V_1,V_2\dots , V_{32}$} + \underbrace{8}_\text{output $X$} + \underbrace{32\cdot6}_\text{output $P_1,P_2\cdot , P_{32}$} = 1224$ input/output blocks for E3M2; 
    \item $\underbrace{32\cdot32}_\text{inputs of $V_1,V_2\dots , V_{32}$} + \underbrace{8}_\text{output $X$} + \underbrace{32\cdot6}_\text{output $P_1,P_2\cdot , P_{32}$} = 1224$ input/output blocks for E2M3; 
    \item $\underbrace{32\cdot32}_\text{inputs of $V_1,V_2\dots , V_{32}$} + \underbrace{8}_\text{output $X$} + \underbrace{32\cdot4}_\text{output $P_1,P_2\cdot , P_{32}$} = 1160$ input/output blocks for E2M1;
    \item $\underbrace{32\cdot32}_\text{inputs of $V_1,V_2\dots , V_{32}$} + \underbrace{8}_\text{output $X$} + \underbrace{32\cdot4}_\text{output $P_1,P_2\cdot , P_{32}$} = 1288$ input/output blocks for INT8.
\end{itemize}
The implementation design was conducted on Virtex UltraScale (xcvu440-flga2892-1-i), which is equipped with 1456 input/output blocks. Table~\ref{res} provides the results of the Xilinx Vivado 2019.1. The proposed design is based on latches implementation (elements with feed-back wires), hence the critical path evaluation is meaningless. The experimental results include the place and routing process.

The architecture of the proposed converter of 32 FP32 numbers consists of three sort of blocks. The first step calculates the greater exponent among 32 FP32 inputs and it costs about 55\% of all hardware resources. The second step transforms the greater exponent to the shared exponent $"X"$ and it costs up to 1\%. The third step generates private elements $P_1,P_2,\dots,P_{32}$ and costs around 44\% of hardware costs.

\setlength{\tabcolsep}{0.2em}
\renewcommand{\arraystretch}{2}

\begin{table}
\centering
\normalsize
\caption{Results of the implementation Virtex UltraScale (xcvu440-flga2892-1-i)}\label{res}
\begin{tabular}{|c||c|c|c|c|c|c|}
\hline
MX-format & E5M2 & E4M3 & E3M2 & E2M3 & E2M1 & INT8 \\
\hline
\hline
LUTs               & 2319 & 2776 & 2230 & 2039 & 1896 & 1614 \\ \hline

critical path (ns) & 51.4 & 80.2 & 52.8 & 62.6 & 65.2 & 50.6 \\
\hline
\end{tabular}
\end{table}

\section{Conclusion}
This paper presents the hardware architecture of a converter that transforms 32 single-precision floating-point numbers (FP32) into six types of MX-format. FPGA implementation results are provided for the six MX-format types: E5M2, E4M3, E3M2, E2M3, E2M1, and INT8. The proposed converter architecture utilizes a three-step algorithm and is implemented using combinational logic.

\section*{Acknowledgment}
The authors gratitude to Ritchie Zhao for his valuable assistance in explaining the rounding and quantization of numbers in the MX-format.



\end{document}